\documentclass[10pt,twocolumn]{article}

\usepackage[margin=0.75in]{geometry}
\usepackage[T1]{fontenc}
\usepackage[utf8]{inputenc}
\usepackage{booktabs}
\usepackage{tabularx}
\usepackage{array}
\usepackage{amsmath}
\usepackage{graphicx}
\usepackage{enumitem}
\usepackage[numbers,sort&compress]{natbib}
\usepackage[hidelinks]{hyperref}
\usepackage{microtype}
\usepackage{placeins}
\usepackage{cuted}
\usepackage{capt-of}

\newcommand{\Description}[1]{}
\newcommand{\StudyEpochs}{12}
\newcommand{\StudySeedCount}{3}
\newcommand{\TotalCases}{89}
\newcommand{\TotalSlices}{10423}
\newcommand{\TrainCases}{70}
\newcommand{\ValCases}{9}
\newcommand{\TestCases}{10}
\newcommand{\TrainSlices}{8351}
\newcommand{\ValSlices}{1021}
\newcommand{\TestSlices}{1051}

\newcommand{\MedSegCases}{10}
\newcommand{\LongCiuCases}{1}
\newcommand{\PlethoraCases}{78}
\newcommand{\SliceMixedTrainCaseCount}{89}
\newcommand{\SliceMixedValCaseCount}{88}
\newcommand{\SliceMixedTestCaseCount}{89}
\newcommand{\SliceMixedTrainValOverlap}{88}
\newcommand{\SliceMixedTrainTestOverlap}{89}
\newcommand{\SliceMixedValTestOverlap}{88}

\newcommand{\MainSliceDiceMean}{0.6665}
\newcommand{\MainSliceDiceStd}{0.0067}
\newcommand{\MainSliceIoUMean}{0.5031}
\newcommand{\MainSliceIoUStd}{0.0070}
\newcommand{\MainCaseDiceMean}{0.2066}
\newcommand{\MainCaseDiceStd}{0.0509}
\newcommand{\MainCaseIoUMean}{0.1181}
\newcommand{\MainCaseIoUStd}{0.0313}
\newcommand{\MainDiceAbsDrop}{0.4599}
\newcommand{\MainDiceRelDropPct}{69.00}
\newcommand{\MainIoUAbsDrop}{0.3850}
\newcommand{\MainIoURelDropPct}{76.52}
\newcommand{\TunedSliceDice}{0.6811}
\newcommand{\TunedSliceIoU}{0.5179}
\newcommand{\TunedCaseDice}{0.3490}
\newcommand{\TunedCaseIoU}{0.2149}
\newcommand{\ArchitectureStudyAvailable}{1}
\newcommand{\AblationStudyAvailable}{1}
\newcommand{\PerBreakdownAvailable}{1}
\newcommand{\ArchSliceUnetDiceMean}{0.7098}
\newcommand{\ArchSliceUnetDiceStd}{0.0083}
\newcommand{\ArchCaseUnetDiceMean}{0.2472}
\newcommand{\ArchCaseUnetDiceStd}{0.0249}
\newcommand{\SamplerSliceDiceMean}{0.6916}
\newcommand{\SamplerCaseDiceMean}{0.2309}
\newcommand{\LossSliceDiceMean}{0.6680}
\newcommand{\LossCaseDiceMean}{0.2363}
\newcommand{\ArchitectureRows}{%
Slice-mixed shared FPN & 0.6665 $\pm$ 0.0067 & 0.5031 $\pm$ 0.0070 \tabularnewline%
Slice-mixed shared U-Net & 0.7098 $\pm$ 0.0083 & 0.5555 $\pm$ 0.0090 \tabularnewline%
Case-disjoint shared FPN & 0.2066 $\pm$ 0.0509 & 0.1181 $\pm$ 0.0313 \tabularnewline%
Case-disjoint shared U-Net & 0.2472 $\pm$ 0.0249 & 0.1444 $\pm$ 0.0143 \tabularnewline%
}
\newcommand{\AblationRows}{%
Slice-mixed baseline & 0.6665 $\pm$ 0.0067 \tabularnewline%
Slice-mixed + Rare-FG & 0.6916 $\pm$ 0.0069 \tabularnewline%
Slice-mixed + Lovasz-CE & 0.6680 $\pm$ 0.0084 \tabularnewline%
Case-disjoint baseline & 0.2066 $\pm$ 0.0509 \tabularnewline%
Case-disjoint + Rare-FG & 0.2309 $\pm$ 0.0360 \tabularnewline%
Case-disjoint + Lovasz-CE & 0.2363 $\pm$ 0.0269 \tabularnewline%
}
\newcommand{\PerClassRows}{%
ground glass & 0.6618 $\pm$ 0.0034 & 0.1762 $\pm$ 0.0588 \tabularnewline%
consolidation & 0.5940 $\pm$ 0.0156 & 0.1398 $\pm$ 0.0815 \tabularnewline%
pleural effusion & 0.7437 $\pm$ 0.0015 & 0.3037 $\pm$ 0.0231 \tabularnewline%
}
\newcommand{\PerSourceRows}{%
longciu & 0.3389 $\pm$ 0.0065 & 0.0718 $\pm$ 0.0485 \tabularnewline%
medseg sirm & 0.4212 $\pm$ 0.0066 & 0.2017 $\pm$ 0.0167 \tabularnewline%
plethora & 0.2487 $\pm$ 0.0005 & 0.1013 $\pm$ 0.0077 \tabularnewline%
}

\begin{document}

\twocolumn[
\begin{center}
{\LARGE\bfseries CTSCAN: Evaluation Leakage in Chest CT Segmentation and a Reproducible Patient-Disjoint Benchmark\par}
\vspace{0.75em}
{\large Anton Ivchenko\par}
\vspace{0.25em}
{\normalsize\texttt{toxa.ivchenko@gmail.com}\par}
\end{center}
\vspace{0.9em}
]

\begingroup
\small
\noindent\textbf{Abstract.} Reported chest CT segmentation performance can be
strongly inflated when train and test partitions mix slices from the same
study. We present CTSCAN, a reproducible multi-source chest CT benchmark and
research stack designed to measure what survives under patient-disjoint
evaluation. The current four-class artifact aggregates \TotalCases{} cases from
PleThora, MedSeg SIRM, and LongCIU, and we show that the original slice-PNG
workflow induces near-complete case reuse across train, validation, and test.
Using the \texttt{playground} environment and an Apple-GPU validation
environment, we run a multi-seed protocol sweep with the same FPN plus
EfficientNet-B0 control configuration under slice-mixed and case-disjoint
evaluation. Across \StudySeedCount{} seeds and \StudyEpochs{} epochs per seed,
the slice-mixed protocol reaches \MainSliceDiceMean{} $\pm$
\MainSliceDiceStd{} foreground Dice and \MainSliceIoUMean{} $\pm$
\MainSliceIoUStd{} foreground IoU, whereas the case-disjoint protocol reaches
\MainCaseDiceMean{} $\pm$ \MainCaseDiceStd{} Dice and \MainCaseIoUMean{} $\pm$
\MainCaseIoUStd{} IoU. Removing patient reuse therefore reduces foreground Dice
by \MainDiceAbsDrop{} absolute (\MainDiceRelDropPct{}\% relative) and
foreground IoU by \MainIoUAbsDrop{} absolute (\MainIoURelDropPct{}\% relative).
CTSCAN packages the corrected benchmark with deterministic split manifests,
explicit weak-supervision controls, a scripted multi-seed protocol sweep, and
reproducible figure generation, providing a reusable basis for
patient-disjoint chest CT evaluation.
\par
\endgroup
\vspace{0.6em}

\section{Introduction}

Chest CT segmentation pipelines are often implemented on top of per-slice PNG
exports from volumetric studies. That design is convenient, but it creates a
serious evaluation risk: if slices are partitioned directly, train and test
sets can share patients even when filenames differ. In that regime, reported
metrics reflect patient reuse as much as they reflect model quality.

This paper studies that problem through CTSCAN, a public multi-source chest CT
segmentation benchmark designed to make patient-disjoint evaluation the default
reproducible path. The central scientific question is not whether one
architecture can be tuned to a strong number. It is whether the evaluation
protocol changes the conclusion about what actually generalizes.

Our thesis is that a large fraction of apparent chest CT segmentation
performance can be explained by evaluation leakage, and that the right
benchmarking question is what signal remains after enforcing patient-disjoint
evaluation. This is consistent with broader warnings that leakage can change
the apparent scientific conclusion in machine-learning-based research
settings~\citep{kapoor2023leakage}. CTSCAN provides a reproducible way to answer
that question through a deterministic case-disjoint benchmark built from a
public multi-source artifact. We therefore ask three questions:

\begin{enumerate}[leftmargin=1.5em]
\item How much does slice-mixed evaluation inflate reported chest CT
segmentation performance?
\item What performance remains once the same benchmark is evaluated with
patient-disjoint splits?
\item What public artifact is needed to make patient-disjoint evaluation
reproducible, auditable, and easy to rerun?
\end{enumerate}

Our answer is that protocol choice changes the absolute scientific conclusion,
not just the benchmark number. We make three concrete contributions:

\begin{itemize}[leftmargin=1.5em]
\item We expose a strong evaluation-leakage effect in a slice-derived chest CT
segmentation benchmark: partitioning individual PNG pairs rather than case IDs
induces near-complete case reuse across train, validation, and test.
\item We release a deterministic patient-disjoint benchmark protocol built from
public chest CT sources, exact split manifests, overlap auditing, and a public
benchmark page that ranks models only under patient-disjoint evaluation.
\item We pair that benchmark with a scripted multi-seed study that runs matched
controls across evaluation protocols, learning curves, and controlled training
variants so the leakage claim is supported by executable evidence rather than by
one-off notebook results.
\end{itemize}

\section{Benchmark Setting}

CTSCAN is a public multi-source chest CT segmentation benchmark implemented as
a slice-derived training and evaluation stack. Each volumetric case is exported
to aligned grayscale PNG slices and semantic masks, making the benchmark easy
to reproduce but also easy to split incorrectly if patient identity is ignored.

This paper studies the four-class public release used for the main benchmark:
background, ground glass, consolidation, and pleural effusion. The same project
also contains a broader eight-class pipeline, but the scientific claim in this
paper is about evaluation protocol rather than ontology breadth. We therefore
hold the training recipe fixed and compare two protocols:

\begin{itemize}[leftmargin=1.5em]
\item \textbf{Slice-mixed}: train, validation, and test are formed directly over
individual PNG pairs.
\item \textbf{Patient-disjoint}: slices are grouped by case before
partitioning, and overlap is explicitly audited.
\end{itemize}

The key question is which performance survives once the split is made
patient-disjoint.

\section{Benchmark Dataset and Protocols}

\subsection{Data composition}

The current four-class PNG benchmark contains \TotalCases{} cases drawn from
three public sources: \PlethoraCases{} PleThora cases, \MedSegCases{} MedSeg
SIRM cases, and \LongCiuCases{} LongCIU case. The benchmark contains
\TotalSlices{} slices. Its label distribution is extremely
imbalanced, as shown in Table~\ref{tab:pixelfreq}.

\begin{table}[t]
\centering
\caption{Pixel-frequency imbalance in the four-class PNG benchmark.}
\label{tab:pixelfreq}
\begin{tabular}{lr}
\toprule
Class & Pixel frequency \\
\midrule
Background & 0.996341 \\
Class 1 & 0.001373 \\
Class 2 & 0.000355 \\
Class 3 & 0.001931 \\
\bottomrule
\end{tabular}
\end{table}

Foreground is therefore sparse, and the pleural-effusion-heavy class dominates
among positive labels.

\subsection{Best slice-mixed reference}

The strongest single-run slice-mixed reference on the four-class benchmark uses an
FPN~\citep{lin2017fpn} with an EfficientNet-B0 encoder~\citep{tan2019efficientnet},
Lov{\'a}sz-Softmax plus cross-entropy loss~\citep{berman2018lovasz}, and a
rare-foreground sampler. Under the slice-mixed protocol, this configuration reaches 0.6811
foreground Dice and 0.5179 foreground IoU. At face value, these look like
strong benchmark results.

\subsection{Protocol comparison}

The reference training notebook operates on per-slice PNG pairs rather than
case identifiers. In the notebook, all converted slice pairs are first pooled
into a single list and \texttt{train\_test\_split} is then applied directly to
that slice list. The resulting protocol is therefore slice-mixed rather than
patient-disjoint.
Collapsing slice stems back to case identifiers reveals near-complete case reuse
across train, validation, and test, as shown in Table~\ref{tab:overlap}.

\begin{table}[t]
\centering
\caption{Case reuse implied by the slice-mixed protocol.}
\label{tab:overlap}
\begin{tabular}{lr}
\toprule
Partition summary & Count \\
\midrule
Train cases & \SliceMixedTrainCaseCount{} \\
Validation cases & \SliceMixedValCaseCount{} \\
Test cases & \SliceMixedTestCaseCount{} \\
Train/Test overlap & \SliceMixedTrainTestOverlap{} \\
Train/Validation overlap & \SliceMixedTrainValOverlap{} \\
Validation/Test overlap & \SliceMixedValTestOverlap{} \\
\bottomrule
\end{tabular}
\end{table}

We therefore use a deterministic case-level protocol as the primary evaluation
setting for patient-disjoint generalization.

\section{Benchmark Release}

CTSCAN packages the patient-disjoint benchmark as a reproducible public release
rather than as an undocumented split change.

\subsection{Deterministic patient-disjoint split}

We added a scripted split builder that rebuilds a case-level split from the
PNG slices plus benchmark metadata. The script groups slices by case ID,
preserves source-aware allocation, writes CSV splits understood by the current
trainer, and emits a machine-readable manifest that reports case counts, slice
counts, per-source composition, and explicit case-overlap checks.

The scripted split contains:
\begin{itemize}[leftmargin=1.5em]
\item train: \TrainCases{} cases, \TrainSlices{} slices
\item validation: \ValCases{} cases, \ValSlices{} slices
\item test: \TestCases{} cases, \TestSlices{} slices
\end{itemize}
All pairwise case overlaps are zero.

\subsection{Explicit supervision controls}

The broader dataset builder supports pseudo-labeled sample studies, but the
benchmark release disables them by default. Any weak-supervision run therefore
requires an explicit command-line opt-in rather than being silently mixed into
the public benchmark.

\subsection{Reproducible study presets}

The benchmark release includes named presets for the shared-control and tuned
reference settings. This removes the need to reconstruct long hyperparameter
command lines by hand when rerunning the study.

\subsection{Scripted study and public benchmark page}

We added a reproducible study driver that launches matched training runs under
the slice-mixed and case-disjoint protocols, caches per-seed metrics, and
aggregates both final test scores and per-epoch learning curves into a single
machine-readable summary. The same artifacts feed a generated benchmark page
and leaderboard, so the public release ranks models only under patient-disjoint
evaluation while retaining slice-mixed numbers to quantify leakage.

\section{Experimental Protocol}

\subsection{Goal}

Our goal is to isolate the effect of split protocol itself. We therefore train
the same segmentation control configuration under two settings on the same
benchmark: a slice-mixed protocol and a deterministic case-disjoint protocol.

\subsection{Settings}

Our primary comparison uses the same FPN~\citep{lin2017fpn} with an
EfficientNet-B0 encoder~\citep{tan2019efficientnet}, Dice-CE loss, image size
256, batch size 8, and identical optimizer settings under both protocols. We
run this configuration for \StudySeedCount{} random seeds and \StudyEpochs{}
epochs per seed while keeping the training recipe fixed across protocols. We
separately retain the strongest tuned reference artifact as context for what
the benchmark can reach under more aggressive single-run tuning.

\section{Results}

\begin{table*}[t]
\centering
\caption{Multi-seed comparison under a shared FPN control.}
\label{tab:mainresults}
\small
\begin{tabular}{lcccc}
\toprule
Protocol & Seeds & Epochs & FG Dice & FG IoU \\
\midrule
Slice-mixed shared FPN & \StudySeedCount{} & \StudyEpochs{} & \MainSliceDiceMean{} $\pm$ \MainSliceDiceStd{} & \MainSliceIoUMean{} $\pm$ \MainSliceIoUStd{} \\
Case-disjoint shared FPN & \StudySeedCount{} & \StudyEpochs{} & \MainCaseDiceMean{} $\pm$ \MainCaseDiceStd{} & \MainCaseIoUMean{} $\pm$ \MainCaseIoUStd{} \\
\bottomrule
\end{tabular}
\end{table*}

Table~\ref{tab:mainresults} shows the central protocol comparison under a
matched training recipe. When the same FPN-EfficientNet-B0 control is trained
under the case-disjoint protocol instead of the slice-mixed protocol:

\begin{itemize}[leftmargin=1.5em]
\item foreground Dice drops by \MainDiceAbsDrop{} absolute,
\item foreground Dice drops by \MainDiceRelDropPct{}\% relative,
\item foreground IoU drops by \MainIoUAbsDrop{} absolute, and
\item foreground IoU drops by \MainIoURelDropPct{}\% relative.
\end{itemize}

\begin{figure*}[!t]
\centering
\includegraphics[width=0.95\linewidth]{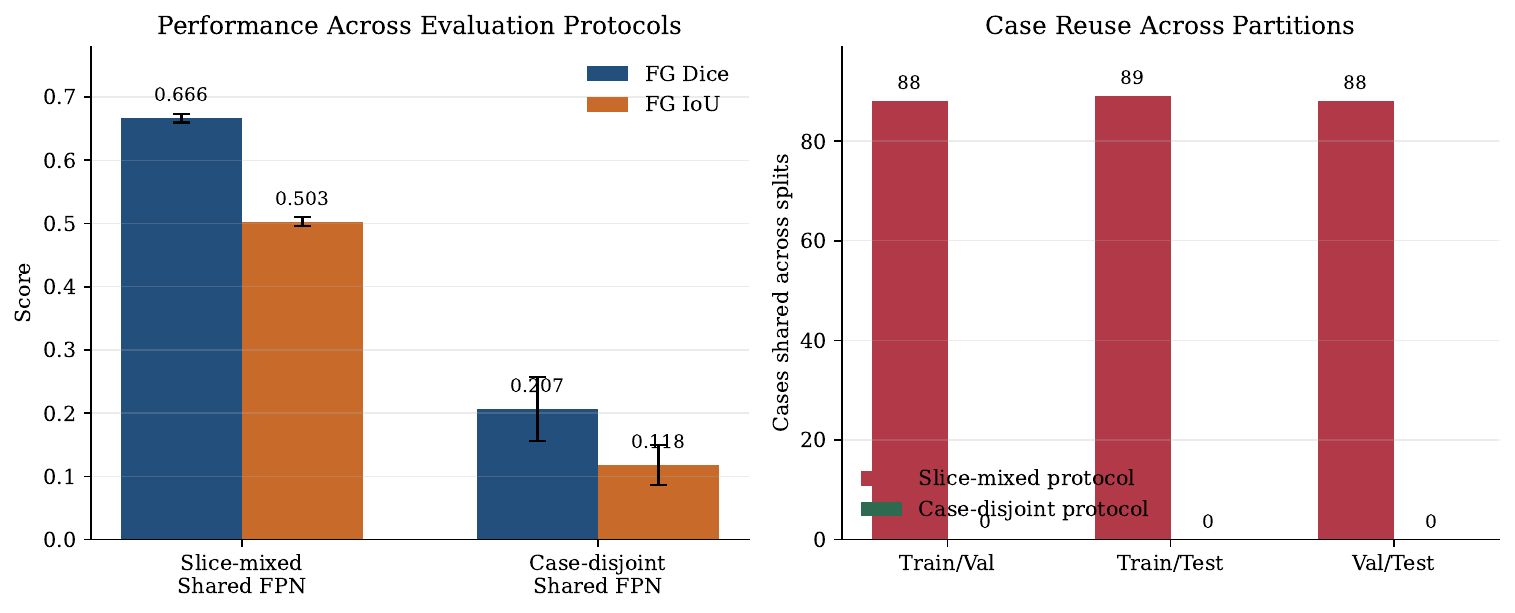}
\caption{Protocol comparison for the shared FPN control. Left: mean foreground
Dice and IoU across seeds under the slice-mixed and case-disjoint protocols.
Right: case reuse counts under the slice-mixed protocol and the case-disjoint
protocol. The figure is generated directly from benchmark metrics and split
artifacts.}
\label{fig:protocolsummary}
\Description{A two-panel summary figure. The left panel is a grouped bar chart
showing foreground Dice and foreground IoU for the slice-mixed and
case-disjoint protocols with error bars across seeds. The right panel is a
grouped bar chart showing case reuse counts for train/validation, train/test,
and validation/test under the slice-mixed and case-disjoint protocols.}
\end{figure*}

\begin{figure*}[t]
\centering
\includegraphics[width=0.95\linewidth]{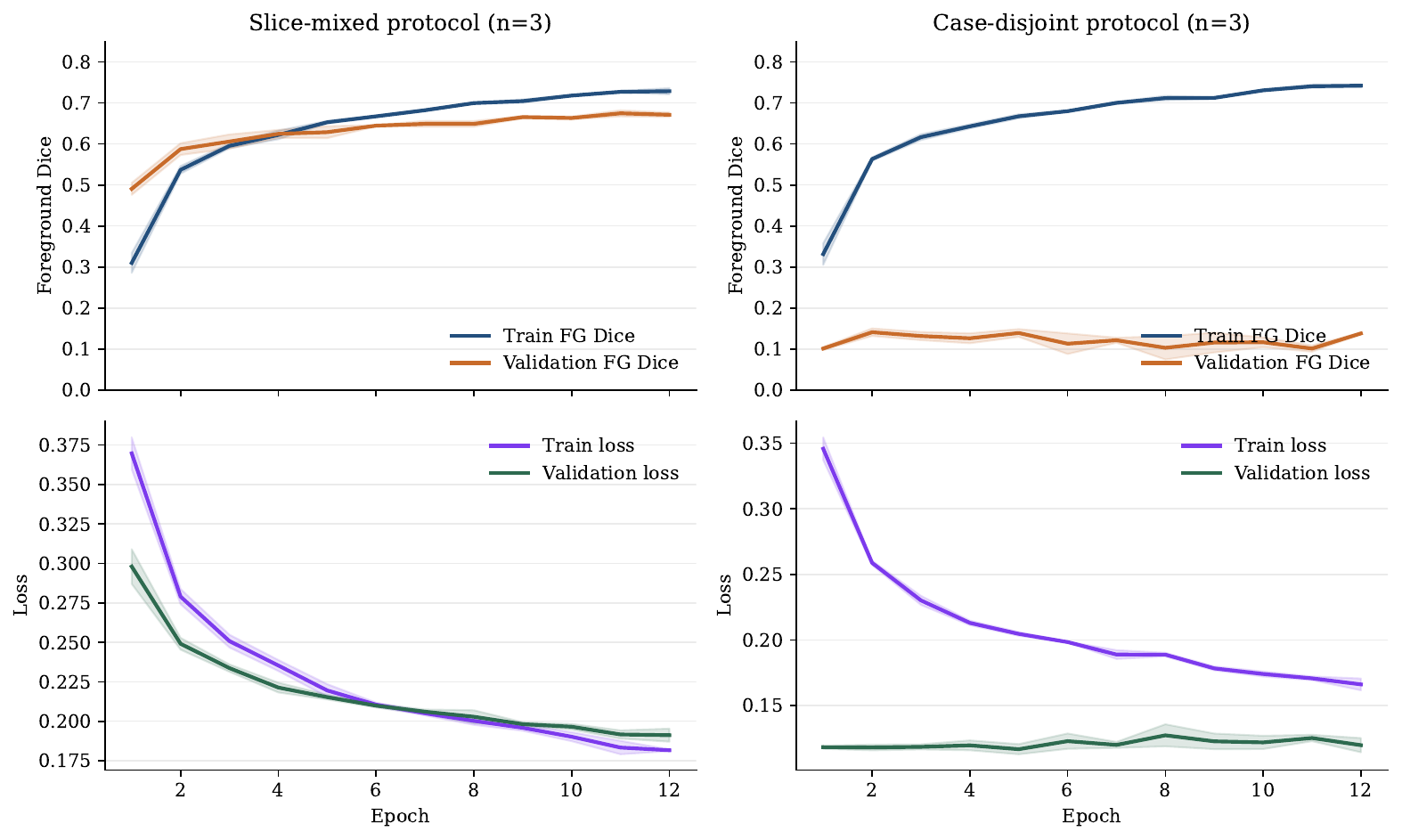}
\caption{Multi-seed training and validation curves for the shared FPN control.
Top: foreground Dice. Bottom: loss. Left: slice-mixed protocol. Right:
case-disjoint protocol. Solid lines show seed means and shaded bands show one
standard deviation.}
\label{fig:protocolcurves}
\Description{A four-panel figure with training and validation curves for the
slice-mixed and case-disjoint protocols. The top row shows foreground Dice and
the bottom row shows loss. Each panel contains mean curves with shaded standard
deviation bands across seeds.}
\end{figure*}

This is a large protocol effect under a controlled, multi-seed experiment. It
shows that split construction materially changes the scientific conclusion even
before aggressive tuning enters the picture. Figure~\ref{fig:protocolsummary}
shows the same gap in the final mean metrics, while
Figure~\ref{fig:protocolcurves} shows that the separation appears early in
training and persists across epochs rather than being driven by a single
outlying run.

As context, the strongest tuned single-run references show
the same direction of effect: the slice-mixed tuned artifact reaches
\TunedSliceDice{} foreground Dice and \TunedSliceIoU{} foreground IoU, whereas
the case-disjoint tuned artifact reaches \TunedCaseDice{} Dice and
\TunedCaseIoU{} IoU. We therefore view the seeded
shared-control comparison as the core evidence and the stronger tuned artifacts
as consistent upper-context points from the same benchmark stack.

Figure~\ref{fig:protocolsummary} summarizes the paper's central message:
protocol choice changes the scientific conclusion about absolute performance,
and the effect is large enough to remain visible after aggregation across seeds.

\ifnum\ArchitectureStudyAvailable>0
\begin{table}[t]
\centering
\caption{Architecture robustness under the same split protocols.}
\label{tab:architecture}
\small
\begin{tabular}{lcc}
\toprule
Setting & FG Dice & FG IoU \\
\midrule
\ArchitectureRows
\bottomrule
\end{tabular}
\end{table}

The leakage effect is not specific to one segmentation family.
Table~\ref{tab:architecture} and Figure~\ref{fig:architecturesummary} show that
the same qualitative collapse appears for a shared U-Net~\citep{ronneberger2015unet}
control: the
slice-mixed U-Net reaches \ArchSliceUnetDiceMean{} $\pm$
\ArchSliceUnetDiceStd{} foreground Dice, whereas the case-disjoint U-Net reaches
\ArchCaseUnetDiceMean{} $\pm$ \ArchCaseUnetDiceStd{}.

\begin{figure*}[t]
\centering
\includegraphics[width=0.82\linewidth]{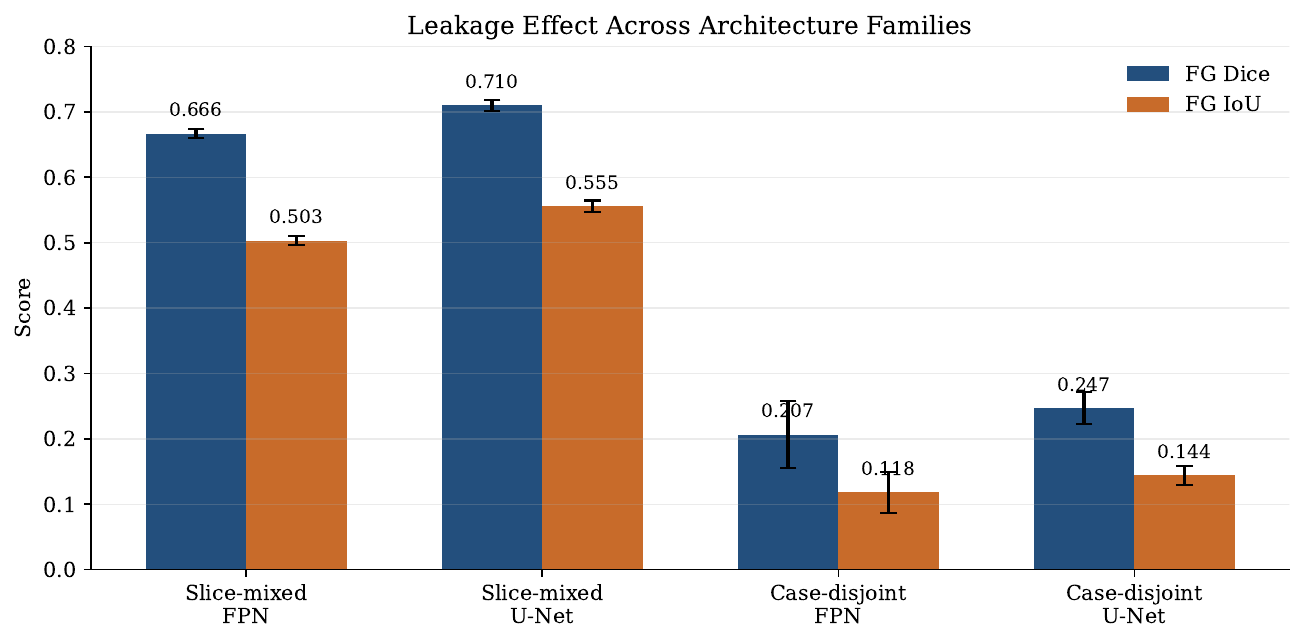}
\caption{Leakage effect across architecture families. Both the shared FPN and
the shared U-Net controls lose substantial foreground accuracy once evaluation
is made patient-disjoint.}
\label{fig:architecturesummary}
\Description{A grouped bar chart comparing foreground Dice and IoU for FPN and
U-Net under slice-mixed and case-disjoint evaluation.}
\end{figure*}
\fi

\ifnum\AblationStudyAvailable>0
\begin{table}[t]
\centering
\caption{Ablations under slice-mixed and case-disjoint evaluation.}
\label{tab:ablations}
\small
\begin{tabular}{lc}
\toprule
Setting & FG Dice \\
\midrule
\AblationRows
\bottomrule
\end{tabular}
\end{table}

Sampler and loss ablations do not erase the protocol effect. Under the
slice-mixed protocol, Rare-FG sampling reaches \SamplerSliceDiceMean{} foreground
Dice and Lovasz-CE reaches \LossSliceDiceMean{}. Under the patient-disjoint
protocol, the corresponding runs reach \SamplerCaseDiceMean{} and
\LossCaseDiceMean{}, respectively. The protocol shift therefore dominates
second-order improvements from training tweaks.

\begin{figure*}[t]
\centering
\includegraphics[width=0.82\linewidth]{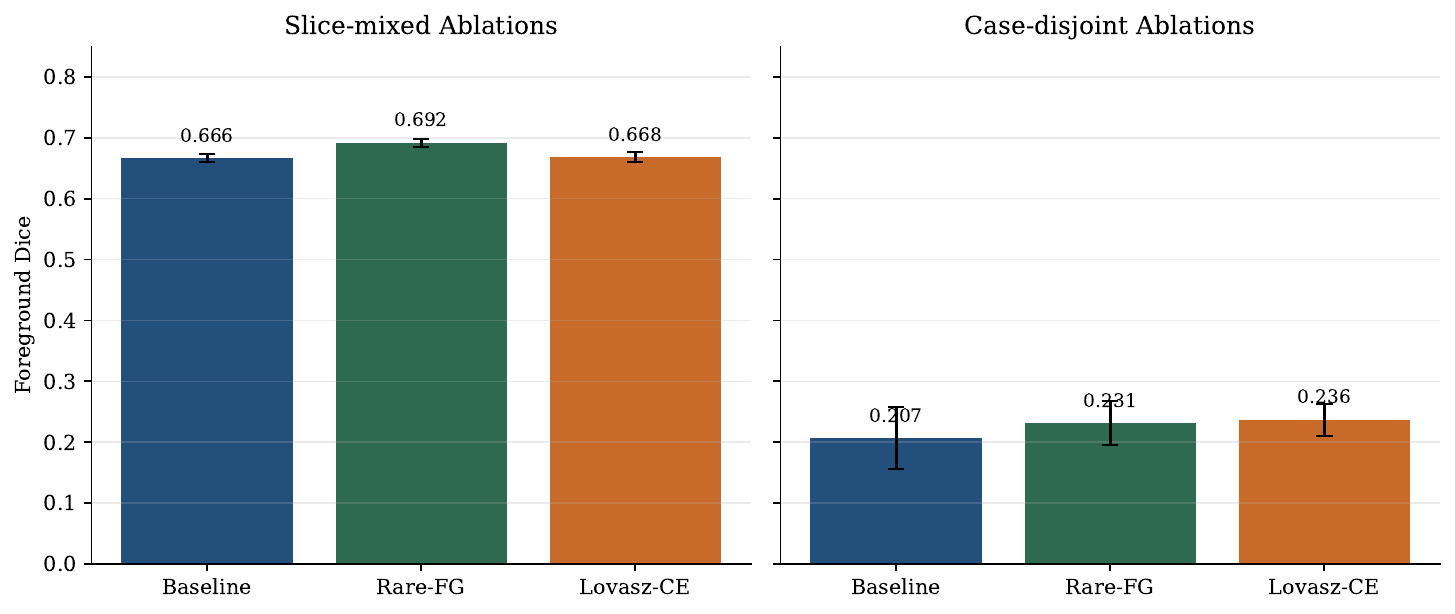}
\caption{Sampler and loss ablations under slice-mixed and patient-disjoint
evaluation. Training tweaks move the score within each protocol, but they do
not close the protocol gap.}
\label{fig:ablationsummary}
\Description{Two bar charts showing foreground Dice for baseline, Rare-FG, and
Lovasz-CE settings under slice-mixed and case-disjoint evaluation.}
\end{figure*}
\fi

\ifnum\PerBreakdownAvailable>0
\begin{table}[t]
\centering
\caption{Per-class foreground Dice for the shared FPN control.}
\label{tab:perclass}
\small
\begin{tabular}{lcc}
\toprule
Class & Slice-mixed Dice & Case-disjoint Dice \\
\midrule
\PerClassRows
\bottomrule
\end{tabular}
\end{table}

\begin{table}[t]
\centering
\caption{Per-source foreground Dice for the shared FPN control.}
\label{tab:persource}
\small
\begin{tabular}{lcc}
\toprule
Source & Slice-mixed FG Dice & Case-disjoint FG Dice \\
\midrule
\PerSourceRows
\bottomrule
\end{tabular}
\end{table}

\fi

\section{Discussion}

\subsection{Implications for Evaluation}

The main result is not a new architecture. It is a correction to how this
problem should be measured. In this benchmark, slice-mixed evaluation
produces strong-looking metrics because the same cases appear across splits.
Once evaluation is made case-disjoint, the absolute scores drop sharply. The
multi-seed curves reported here show that this is not a one-off run artifact:
the gap persists across seeds and throughout training.

The broader implication is that chest CT segmentation results should not be read
without asking how partitions were constructed. For slice-derived datasets,
patient-disjoint splitting is not a cosmetic cleanup step. It determines whether
the benchmark measures memorization of patient-specific appearance or genuine
cross-patient generalization. Because this slice-level split pattern is easy to
reproduce in related chest CT training code, we publish this artifact as a
deliberate corrective: the benchmark, manifests, and sweep script make the
patient-disjoint protocol the default executable path rather than an afterthought.

\subsection{What CTSCAN Provides}

CTSCAN contributes a reproducible evaluation artifact around that point:
\begin{itemize}[leftmargin=1.5em]
\item a multi-source chest CT benchmark construction path,
\item deterministic case-disjoint split generation with explicit overlap
auditing, and
\item a pretrained-backbone segmentation testbed with a scripted multi-seed
protocol comparison workflow.
\end{itemize}

\section{Experimental Scope}

This study establishes the leakage effect clearly, but it does not exhaust the
full space of benchmark development.

\begin{enumerate}[leftmargin=1.5em]
\item The current public release covers \TotalCases{} cases and is dominated by
PleThora, so broader cross-source coverage remains important.
\item The present benchmark focuses on the four-class PNG artifact; extending
the same patient-disjoint discipline to the broader eight-class pipeline is the
next benchmark-expansion step.
\item A fully external held-out site benchmark would strengthen the distribution
shift story further.
\end{enumerate}

\section{Artifact and Verification Checklist}

The paper-facing artifacts created during this work include:

\begin{itemize}[leftmargin=1.5em]
\item a deterministic case-split builder,
\item a scripted multi-seed protocol sweep with aggregate JSON and Markdown
summaries,
\item a reproducible figure-generation script driven by benchmark artifacts,
\item a generated public benchmark page with machine-readable leaderboard
artifacts,
\item tests for the split builder and study tooling,
\item a machine-readable split manifest,
\item multi-seed split-comparison metrics and learning curves, and
\item case-disjoint and slice-mixed trainer presets for paper runs.
\end{itemize}

The public repository for the paper scripts and release packaging is
\url{https://github.com/TonyIvchenko/playground/tree/main/src/ctscan}.

Verification coverage for the benchmarked paths is strong:

\begin{itemize}[leftmargin=1.5em]
\item full CTSCAN test suite in \texttt{playground}: 168 passed,
\item CTSCAN study tests in \texttt{playground}: 7 passed,
\item CTSCAN study tests in the disposable validation environment: 7 passed.
\end{itemize}

\section{Conclusion}

This paper shows that evaluation leakage materially changes the measured
conclusion in chest CT segmentation. In the CTSCAN benchmark release, the
original slice-mixed workflow induces near-complete case reuse, and a
multi-seed shared-control experiment still reports \MainSliceDiceMean{} $\pm$
\MainSliceDiceStd{} foreground Dice and \MainSliceIoUMean{} $\pm$
\MainSliceIoUStd{} foreground IoU under that slice-mixed regime. Under a
deterministic case-disjoint benchmark built from the same artifact, the same
control reaches \MainCaseDiceMean{} $\pm$ \MainCaseDiceStd{} Dice and
\MainCaseIoUMean{} $\pm$ \MainCaseIoUStd{} IoU. CTSCAN therefore contributes
more than another
training script: it provides a reproducible patient-disjoint benchmark and
seeded evaluation workflow showing what performance survives once leakage is
removed.

\FloatBarrier
\begin{strip}
\centering
\vspace{-0.8\baselineskip}
\includegraphics[width=0.9\textwidth]{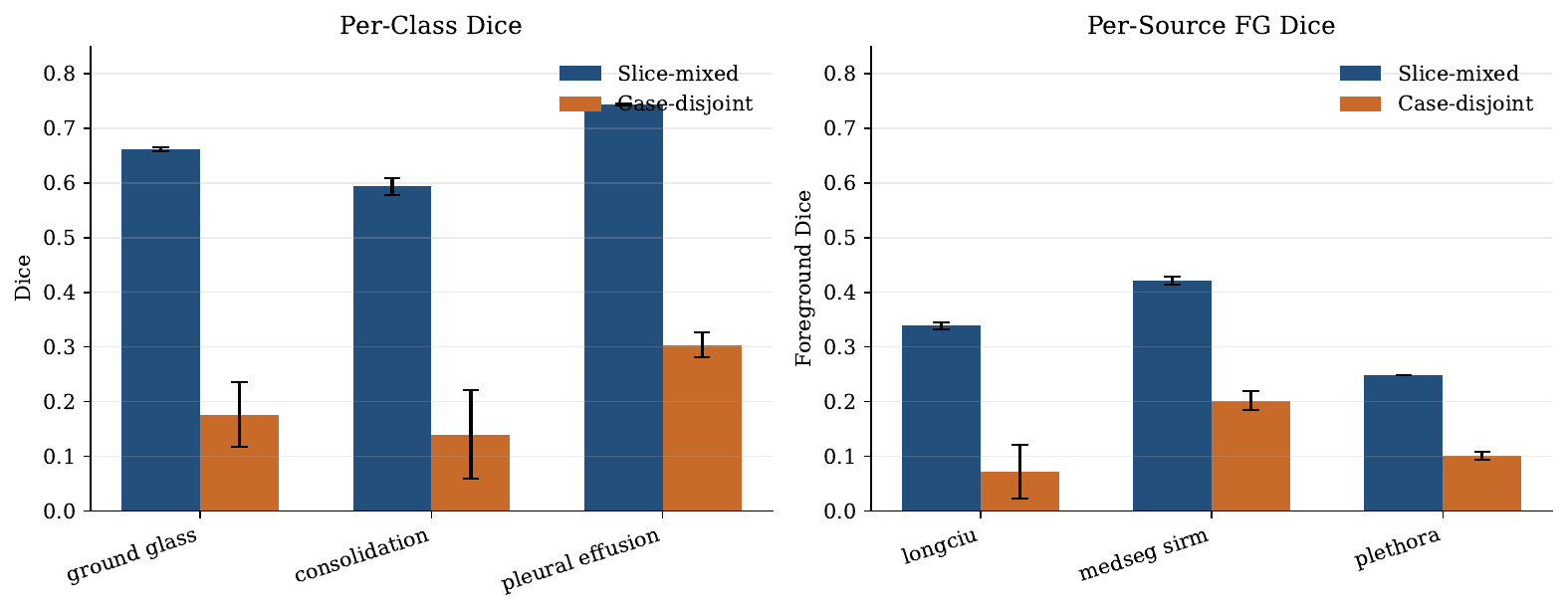}
\captionof{figure}{Per-class and per-source breakdown under the shared FPN control. The same split change that inflates the aggregate score also alters the apparent difficulty of individual classes and data sources.}
\label{fig:breakdownsummary}
\end{strip}
\vspace{-0.5\baselineskip}

\bibliographystyle{abbrvnat}
\bibliography{references}

\end{document}